\documentstyle[12pt]{article}
\begin{document}
\newcommand{\bea}{\begin{eqnarray}}
\newcommand{\eea}{\end{eqnarray}}
\newcommand{\be}{\begin{equation}}
\newcommand{\ee}{\end{equation}}
\newcommand{\non}{\nonumber}
\newcommand{\ov}{\overline}
\global\parskip 6pt
\begin{titlepage}
\begin{center}
{\Large\bf On Dijkgraaf-Witten Type Invariants }\\
\vskip .50in
Danny Birmingham \footnote{Supported by Stichting voor Fundamenteel
Onderzoek der Materie (FOM)\\
Email: Dannyb@phys.uva.nl}     \\
\vskip .10in
{\em Universiteit van Amsterdam, Instituut voor Theoretische Fysica,\\
Valckenierstraat 65, 1018 XE Amsterdam, \\
The Netherlands} \\
\vskip .50in
Mark Rakowski\footnote{Email: Rakowski@maths.tcd.ie}   \\
\vskip .10in
{\em School of Mathematics, Trinity College, Dublin 2, Ireland}  \\
\end{center}
\vskip .10in
\begin{abstract}
   We explicitly construct a series of lattice models based upon the
gauge group $Z_{p}$ which have the property of subdivision
invariance, when the coupling parameter is quantized and the field
configurations are restricted to satisfy a type of mod-$p$ flatness
condition. The simplest model of this type yields the Dijkgraaf-Witten
invariant of a $3$-manifold and is based upon a single link,
or $1$-simplex, field. Depending upon the manifold's dimension,
other models may have more than one species of field variable, and
these may be based on higher dimensional simplices.
\end{abstract}
\vskip .5in
\begin{center}
ITFA-94-07\\
February 1994
\end{center}
\end{titlepage}

\section{Introduction}

An intriguing three dimensional lattice model was constructed by
Dijkgraaf and Witten in \cite{DW}. By general considerations in
gauge theory, it was shown that three dimensional Chern-Simons
theories are classified by the cohomology classes in $H^{4}(BG,Z)$,
where $BG$ is the universal classifying space for the group $G$.
In the case of a finite group, they showed that the Boltzmann
weight of such a theory was a 3-cocycle in $H^{3}(BG,R/Z)$; the
cocycle condition being equivalent to the equation which guaranteed
subdivision invariance of the lattice model. Subdivision invariance
is, roughly speaking, the analogue of metric independence of a
continuum theory.

In this paper, we will find a more concrete formulation
for lattice models which have some features similar to the
Dijkgraaf-Witten theory; their theory will appear as the simplest
example. Extensions of that model to all odd dimensions,
which was implicit in their formulation, appear as one series of
models in our construction. The Chern-Simons type series
just mentioned is based
on dynamical variables associated only to links of the lattice,
and is the closest to standard gauge theory. We also find other
theories in our approach which have a superficial resemblance to the
continuum $U(1)$ theory introduced by Schwarz \cite{AS}, which was
related to Ray-Singer, and equivalently, Franz-Reidemeister, torsion.
These theories will also involve lattice variables associated
to higher dimensional simplices. Additional models which do not really
lie within either of these two categories will also be formulated.
Generically, this construction falls
outside of the scope of \cite{DW} which is rooted in link based
gauge theory.

We work exclusively with the gauge group $Z_{p}$. Subdivision invariance
follows naturally in each model when the field configurations are
restricted to satisfy a type of mod-$p$ flatness condition. While in
three dimensions subdivision invariance of the partition function
is sufficient to conclude that one has a topological invariant, the
situation is more delicate in higher dimensions. There, subdivision
invariance yields a combinatorial invariant of the piecewise
linear structure. This situation is analogous to the continuum model
phenomenon where metric independence allows one to conclude
immediately that one has a diffeomorphism invariant, though further
considerations may show that the theory is topological.

\section{General Formalism}

   A lattice model is based on a simplicial complex which combinatorially
encodes the topological structure of some manifold.
Let us recall some of the essential ingredients that are required
in such a formulation; we refer the reader to \cite{JM,Rot,Still}
for a more complete account.

   Let $V = \{ v_{i} \}$ denote a finite set of $N_{0}$ points which we will
refer to as the vertices of a simplicial complex. An ordered $k$-simplex
is an array of $k+1$ distinct vertices which we denote by,
\bea
[ v_{0},\cdots ,v_{k} ] \;\;.
\eea
It will usually be convenient to use simply the indices themselves
to label a given vertex when no confusion will arise, so the above
simplex is denoted more economically by $[0,\cdots ,k]$.
Pictorially, a $k$-simplex should be regarded as a point,  line segment,
triangle, or tetrahedron for $k$ equals zero through three respectively.
A simplex which is spanned by any subset of the vertices is called a face
of the original simplex.
An  orientation of a simplex is a choice of ordering of its vertices,
where we identify orderings that differ by an even permutation, but for
the models described here we will require an ordering of all vertices.
One then checks that the invariant we compute is actually independent
of the choice made in vertex ordering.

The boundary operator $\partial$ on the ordered simplex
$\sigma=[v_{0},\cdots,v_{k}]$  is defined by,
\bea
\partial \,\sigma = \sum^{k}_{i=0}\; (-1)^{i}\, [v_{0},\cdots,
\hat{v}_{i},\cdots,v_{k}]\;\;,
\eea
where the `hat' indicates a vertex which has been omitted. It
is easy to show that the composition of boundary operators is zero;
$\partial^{2} = 0$.

We model a closed $n$-dimensional manifold as a collection
$K= \{ \sigma_{i} \}$ of $n$-simplices
constructed from the set of vertices $V$, subject to a few technical
conditions. Most importantly, every $(n-1)$-face of any given $n$-simplex
appears as an $(n-1)$-face of precisely two different $n$-simplices in
the collection $K$. One thinks of the $n$-simplices then as glued
together along $(n-1)$-faces.
There is an additional restriction on the
``link'' of a vertex for the the simplicial complex to represent
a manifold, but this condition will not play
a role in the sequel and we refer the reader to \cite{Still} for a
more complete discussion.

The dynamical variables in the theories we construct will be
objects which assign an element in the cyclic group $Z_{p} = Z / pZ$,
which we represent as the set of integers,
\bea
\{ 0,\cdots , p-1 \} \;\; ,
\eea
to ordered simplices of some specified dimension. We call these dynamical
variables $k$-colours with coefficients in $Z_{p}$, and denote
the evaluation of some $k$-colour $B^{(k)}$ on the ordered $k$-simplex
$[0,\cdots ,k]$ by
\bea
< B^{(k)}, [0,\cdots ,k] > = B_{0 \cdots k} \in Z_{p} \;\; .
\eea
The superscript $(k)$ will usually be omitted when its value is clear
from context.
It is important to note that we are assigning a $Z_{p}$
element in a way which depends on the ordering of vertices in the simplex;
we do not have the rule $B^{(1)}_{01} = - B^{(1)}_{10}$, for example.
Instead, we shall assume that,
\bea
B^{(1)}_{10} = - B^{(1)}_{01} \;\;mod\;\; p \;\; ,
\eea
and similarly extend this to a $k$-colour for odd permutations of the
vertices. The case closest to conventional lattice gauge theory is
where a 1-colour variable is assigned to every 1-simplex in the complex.

The coboundary operator $\delta$ acts on the dynamical variables as follows.
Given a $(k-1)$-colour, an application of the coboundary operator
produces an integer in $Z$, when evaluated on an ordered $k$-simplex, namely
\bea
< \delta B^{(k-1)} , [0,\cdots , k] > &=& < B, \partial [0,\cdots , k]>
\nonumber \\
&=& B_{123\cdots k} - B_{023 \cdots k} + B_{013\cdots k} - \cdots \;\; .
\eea
We must emphasize that the above sum of integers is not taken with
modular $p$ arithmetic; it is simply an element in $Z$.
In cases where we will need to take some combination mod-$p$,
we will put those terms between square brackets, so for example,
\bea
[ a + b ] = a + b \;\; mod\,\, p \;\; .
\eea

There is also a cup product operation on colours which takes a $k$-colour
$B^{(k)}$ and a $l$-colour $C^{(l)}$ and gives an integer in $Z$
when evaluated on a $(k+l)$-ordered simplex:
\bea
< B\cup C, [0,\cdots ,k+l ] > = B_{0\cdots k} \cdot C_{k\cdots k+l} \;\; .
\eea
Note once again that this product is in $Z$ and the value is not taken
mod-$p$.

Let us now put these ingredients together and define our theories. First,
we must be given some oriented simplicial complex $K$ which we take to
represent a manifold of dimension $n$. One then has some collection
of $n$-simplices defined up to orientation. Take the vertex set of
this complex and give it an ordering. This is done arbitrarily and we
will have to show that our construction is independent of this choice.
Now we can write down an ordered collection of the $n$-simplices; each
of the simplices is written in ascending order and a sign in front of that
simplex indicates whether that ordering is positively or negatively
oriented with respect to the orientation of the complex K. Let us denote
this ordered set of $n$-simplices by $K^{n}$,
\bea
K^{n} = \sum_{i} \;\; \epsilon_{i}\, \sigma_{i}\;\; ,
\eea
where the index $i$ runs over the ordered $n$-simplices $\sigma_{i}$ and
$\epsilon_{i}$ is a sign which indicates the orientation. We will
assign a Boltzmann weight $W[K^{n}]$ to $K^{n}$ by taking a
product of factors,
one for every $n$-simplex,
\bea
W[K^{n}] = \prod_{i} \;\; W[\sigma_{i}]^{\epsilon_{i}}\;\; .
\eea
Each of the individual factors is a nonzero complex number and will be some
function of the colours. The details of which colours we use and how the
function is defined will depend on the particular model. Finally, the
partition function, which we will require to be a
combinatorial invariant,
is defined to be a quantity which is proportional to the sum
of the Boltzmann weights over all colourings,
\bea
Z = \frac{1}{|G|^{N}}\sum_{colours} \;\; W[K^{n}] \;\; .\label{z}
\eea
Here $N$ is the total number of colour summations and $|G|$ is the order
of the group where the colours take their values. In a theory based
entirely on a single 1-colour field, for example, this number is
equal to the number of 1-simplices in the complex. This factor simply
serves to normalize all the group summations to have unit volume.
Let us make all of this very explicit by defining some specific models.

\section{The Dijkgraaf-Witten Invariant}

The simplest model of the type we are describing will lead to the
Dijkgraaf-Witten invariant of $3$-manifolds \cite{DW}. Further analysis
of this model has been presented in \cite{AC1}-\cite{FQ}.
So, let us be given a
simplicial complex of dimension 3 and an ordering of vertices
as described above.
This model will be constructed out of a single $1$-colour (with values
in $Z_{p}$) denoted by $A$. The weight assigned to some ordered
$3$-simplex $[0,1,2,3]$ is:
\bea
W[[0,1,2,3]]&=& \exp \{ \beta\, < A\cup \delta A, [0,1,2,3] > \}  \\
&=& \exp \{ \beta \, A_{01}\, (A_{12} + A_{23} - A_{13}) \}
\;\; .\nonumber
\eea
Here $\beta$ is a complex number which at this stage is unrestricted.
Clearly, our motivation for taking this particular structure is to
try and mimic the action of a continuum Chern-Simons theory.
We will now see that the requirement of subdivision invariance will
quantize this coupling parameter.

Consider the subdivision of a specific ordered $3$-simplex $[0,1,2,3]$
obtained by installing a new vertex $c$ at the center and linking it
to the other 4 vertices; symbolically,
\bea
[0,1,2,3] \rightarrow [c,1,2,3] - [c,0,2,3] + [c,0,1,3] - [c,0,1,2]\;\;.
\label{sd}
\eea
Let us declare this new vertex to be the first in the total ordering
of all vertices. It is a simple exercise to show that,
\bea
& & W[[0,1,2,3]\; \exp\{- \beta\, <\delta A \cup \delta A,
[c,0,1,2,3]>\} \label{w5} \\
&=& W[[c,1,2,3]]\, W[[c,0,2,3]]^{-1} W[[c,0,1,3]] \,
W[[c,0,1,2]]^{-1} \;\; .\nonumber
\eea
Thus, we see that our Boltzmann weight is not generally invariant under
the replacement of the original Boltzmann factor of $W[[0,1,2,3]]$ by
the 4 factors on the right hand side of (\ref{w5}); there is this
added ``insertion'' which somehow must be trivialized. While one might
imagine other more complicated suggestions, the conditions that lead
to the Dijkgraaf-Witten invariant are to impose a restriction on the
sum over colourings and on the parameter $\beta$. Those conditions
are to take $s = e^{\beta}$ to be a $p^{2}$ root of unity ($s^{p^{2}}=1$)
and to restrict
the sum over colourings to those which satisfy
\bea
\delta \, A = 0 \;\; mod\;\; p \;\;,
\eea
for all $2$-simplices in the complex $K$. This restriction shall be
termed a ``flatness" condition; for example
on the $2$-simplex $[0,1,2]$, we have the restriction
\be
[ A_{01} + A_{12} - A_{02}] = 0  \;\;.
\ee
We remind the reader that the brackets denote a sum which is to be
taken mod-$p$, so this particular equation can also be written as
\be
[ A_{01} + A_{12} ] = A_{02} \;\; .
\ee
With only these flat field configurations,
the product $\delta A \cup \delta A$ is clearly a multiple of $p^2$
and the above insertion becomes unity. The resulting identity (\ref{w5})
shall be referred to as the $5W$ identity. It should be remarked that
subdivision invariance is achieved without the necessity of summing over
the additional colour fields attached to the vertex $c$, and this will be
a general feature of the models presented here.
Notice also that the Boltzmann weight of $[0,1,2,3]$ becomes
\bea
\exp\{ \frac{2\pi i k}{p^{2}} \, A_{01} ( A_{12}
+ A_{23} - [A_{12} + A_{23}] )\} \;\; ,
\eea
with $k\in\{0,\cdots,p-1\}$. This is precisely the well known representation
of a $3$-cocycle for the group cohomology of $Z_{p}$ with coefficients
in $Z_{p}$ (or $U(1)$).

As discussed in \cite{DW}, one can now check that the Boltzmann
weight is gauge invariant for a closed manifold. This property,
together with a verification that the partition function is independent
of the chosen vertex ordering, follows immediately from the $5W$
identity.

\section{Another Model in Three Dimensions}

Having illuminated the general formalism, which in the case of a single
$1$-colour yields the Dijkgraaf-Witten model, we can immediately
consider generalizations. In three dimensions, we have the obvious choice
of a theory with two independent $1$-colour fields.  Let us now treat this
theory is some detail. The Boltzmann weight
of an ordered $3$-simplex $[0,1,2,3]$ is defined as:
\bea
W[[0,1,2,3]] &=& s^{<B\,\cup \,\delta A,\,[0,1,2,3]>} \nonumber\\
&=& s^{B_{01}\,(A_{12} \,+\, A_{23}\, -\, A_{13})}  \;\;,
\label{b1}
\eea
where the two independent $1$-colour fields are denoted by
$B$ and $A$.

Our first duty is to consider the behaviour of the theory under
the subdivision of eqn. (\ref{sd}), and we find
\bea
& &W[[0,1,2,3]]\,s^{- <\delta B \,\cup \,\delta A,\, [c,0,1,2,3]>} \non\\
&=&W[[c,1,2,3]]\,W[[c,0,2,3]]^{-1}\,W[[c,0,1,3]]\,W[[c,0,1,2]]^{-1} \;\;.
\label{b2}
\eea
In this case, we see that invariance under subdivision can be achieved
by again quantizing the coupling scale $s$ to be a $p^{2}$ root of unity,
and restricting the sum over colourings to those which satisfy the
``flatness" conditions:
\bea
[ \delta B ] = [ \delta A ] = 0 \;\; . \label{b3}
\eea
The Boltzmann weight of a single ordered $3$-simplex then assumes the form
\be
W[[0,1,2,3]] = \exp\{\frac{2\pi i k}{p^{2}}\,
B_{01}\,(A_{12} \,+\, A_{23}
\,-\, [A_{12} + A_{23}] )\}  \;\;,      \label{b4}
\ee
where $k \in \{0,\cdots ,p-1 \}$ as before.

Let us now address the issue of gauge invariance on closed manifolds.
We wish to show that the Boltzmann weight (\ref{b4}) is invariant under
independent gauge transformations of the $A$ and $B$ colour fields.
Consider the $A$ and $B$ colour fields defined on the ordered $1$-simplex
$[0,1]$; then, the gauge transformations of those fields are defined as
\bea
A'_{01} &=& [A_{01} + k_{0} - k_{1} ] \;\;,  \non\\
B'_{01} &=& [B_{01} + l_{0} - l_{1} ] \;\;.
\eea
Here, the $k$ and $l$ fields are $0$-colours defined on the vertices
of the complex.
It suffices to consider one example, so let us treat the case of a
gauge transformation of $A$ and $B$ at the $1$-vertex. The
transformed Boltzmann weight, by definition, is given by,
\bea
W'[[0,1,2,3]] &=& s^{[B_{01} - l_{1}]\,([k_{1} + A_{12}] \,+\,
A_{23}
\,-\, [k_{1} + A_{12} + A_{23}])}\non\\
&=& W[[0,1,2,3]]\,s^{B_{01}\,([k_{1} \,+\, A_{12}] \,-\, A_{12}
\,-\, [k_{1} + A_{12} + A_{23}] \,+\, [A_{12} + A_{23}])}\non\\
&.& s^{-\,l_{1}([k_{1} + A_{12}] \,+\, A_{23} \,-\,
[k_{1} + A_{12} + A_{23}])}\;\;.   \label{bgt}
\eea
In order to prove that the Boltzmann weight of a simplicial complex
for a closed manifold is indeed gauge invariant,
we need to show that the additional terms
generated on the right hand side of eqn. (\ref{bgt}) are cancelled by
other factors in the Boltzmann weight of the complex.
To see this, one makes use
of the fact that on a closed complex, each $2$-simplex is common to
precisely two $3$-simplices. It is then a simple matter to check
for the required cancellation.

As we have noted, the Boltzmann weight is defined with an arbitrary
choice of ordering of the vertex set $V$. In order to establish
independence of our results on the choice of ordering, we require a few
identities. Given the Boltzmann weight on an ordered $3$-simplex
$[0,1,2,3]$, it suffices to examine the behaviour under
orientation reversal of the $0-1$, $1-2$, and $2-3$, vertices.
Again, we shall establish that the value of the partition function
is indeed independent of the vertex ordering, on a closed complex.

Under reversal of the $0-1$ vertices, for example, $W[[0,1,2,3]]$ is
replaced by $W[[1,0,2,3]]^{-1}$, and we have the result
\be
W[[1,0,2,3]]^{-1} = W[[0,1,2,3]]\, s^{B_{01}\,([A_{01} + A_{12}]
\,-\,A_{12} \,-\, [A_{01} + A_{12} + A_{23}]
\,+\, [A_{12} + A_{23}])}\;\;.
\ee
Similarly, reversal of the $1-2$ vertices, yields
\bea
W[[0,2,1,3]]^{-1} &=& W[[0,1,2,3]]\, s^{-\,B_{01}(\,A_{12} \,+\,
A_{21})}\non\\
&.& s^{-\,B_{12}(\,A_{21} \,+\, [A_{12} + A_{23}]
\,-\,A_{23})}
\;\;,
\eea
and finally, $2-3$ reversal gives
\be
W[[0,1,3,2]]^{-1} = W[[0,1,2,3]]\, s^{-\, B_{01}(\,A_{23}
\,+\, A_{32})}\;\;.
\ee

To actually establish order independence of the Boltzmann weight
for a closed manifold, we again take recourse to the fact that
each $2$-simplex is common to precisely two $3$-simplices.
Again, we find that the required cancellations occur.

At this point, we have shown that to achieve subdivision invariance, we must
restrict the sum over colourings to those which satisfy the ``flatness"
conditions on each $2$-simplex in the simplicial complex.
Recall that we began with a partition function, (\ref{z}),
which was defined
with respect to a sum over all colourings, and with a normalization
factor of $|G|$ for each independent colour field summation, i.e.,
$|G|^{-2 N_{1}}$, in the theory under study, where $N_{1}$
is the number of $1$-simplices in the complex.
However, the
subdivision invariant Boltzmann weight is one which contains an insertion
of delta functions which impose these flatness restrictions. It only
remains to check the overall scale of the partition function.

This can be obtained
quite easily by noting that the product of delta functions
before and after subdivision are equal, up to the scale factor $|G|^{3}$.
Since the number of $3$-simplices increases by $3$ under this move,
the subdivision invariant partition function contains the
normalization factor, $|G|^{2(N_{3} - N_{1})}$.
One can rewrite this by noting that for the case of a closed $3$-manifold,
the Euler number is zero:
$N_{3} - N_{2} + N_{1} - N_{0} = 0$. Furthermore,
for the case of a closed complex,
we have the restriction that $N_{2} = 2 N_{3}$, and hence the
subdivision invariant partition function can be written as:
\be
Z = \frac{1}{|G|^{2N_{0}}} \sum_{flat}W[K^{n}]\;\;,
\ee
where the sum is now over all flat colourings.

Since the Boltzmann weight and the delta function restrictions are
gauge invariant objects, one has the freedom to gauge fix
arbitrarily the values of a certain number
of the colour configurations.
In the case of a $1$-colour field, the maximal allowable gauge fixing
is called a maximal tree. A simple argument shows that a maximal tree
is specified by the requirement that it should contain no closed
$2$-simplices.
Given the vertex set of $N_{0}$ elements, it is clear that
an ordering exists such that the maximal tree contains $N_{0} -1$
links. In this way, the partition function can be reduced to
a sum over all gauge inequivalent flat colourings (denoted
as $flat '$), with a
normalization as follows:
\be
Z = \frac{1}{|G|^{2}} \sum_{flat '} \;  W[K^{n}] \;\; .
\ee
Therefore, we note that the normalization coincides with that used
in the definition of the Dijkgraaf-Witten theory,
where the partition function is
defined as a sum over all inequivalent flat connections,
$Hom(\pi_{1}(K),G)$.

{}From a practical point of view, the freedom to perform this gauge fixing
facilitates the evaluation of the partition function, to which
we now turn.
For the case of the $3$-sphere, $S^{3}$, a suitable simplicial
complex is provided by the boundary of a single $4$-simplex. An easy
calculation then shows that there is only a single gauge inequivalent
flat colouring, for both the $A$ and $B$ field. The subdivision invariant
value of the partition function is therefore:
\be
Z(S^{3}) = \frac{1}{|G|^{2}}\;\;,
\ee
for all groups $G = Z_{p}$, and all roots of unity $s$.

An equally simple calculation establishes the result,
\be
Z(S^{2} \times S^{1}) = 1\;\;,
\ee
for all $Z_{p}$, and all roots of unity $s$.
Both these results yield the square of the value
obtained in the $Z_{p}$ Dijkgraaf-Witten theory; this will not be the
case in the next example.

An interesting case to consider is provided by the real projective
$3$-space, $RP^{3}$, and we shall deal here with the gauge group $Z_{2}$.
We refer to \cite{tric}, where a convenient
simplicial complex in terms of a small number of vertices is provided.
One should bear in mind, however, that attention must be paid to
the relative orientation of the simplices in the triangulation of
ref. \cite{tric}, so that the boundary of the complex is zero.
The relevant flatness conditions can then be solved, and one finds
that each of the independent 1-colour fields $A$ and $B$ has
2 gauge inequivalent flat solutions. When a non-trivial $4$-th
root of unity is taken for $s$, only
one of the 4 total
field configurations has a Boltzmann weight different from 1,
and the result is,
\be
Z(RP^{3}) = \frac{1}{4} \, ( 1 + 1 + 1 - 1) = \frac{1}{2} \;\;.
\ee
The point to note here is that this value differs from the
calculation in the $Z_{2}$ Dijkgraaf-Witten theory, where
a value of zero is obtained. It is more meaningful, however, to compare
the $B\delta A$ model with the $Z_{2} \times Z_{2}$ Dijkgraaf-Witten
theory. One nontrivial way to represent a  group cocycle in that case
is to take the action to be a sum of two independent Chern-Simons
type terms,
\bea
A \cup \delta A + B \cup \delta B \;\; . \label{cs2}
\eea
The partition function simply factorizes and one merely has to square
the $Z_{p}$ result. Once again a value of
zero is obtained for $RP^{3}$ when a nontrivial 4-th root of
unity is taken for
$s$. However, the $B \delta A$ model we have been discussing has
a Boltzmann weight which can be regarded as a function from
$G \times G \times G$ to $Z_{p}$  (where $G=Z_{p}\times Z_{p}$)
which satisfies the equation for
subdivision invariance. This follows from associating one copy of $Z_{p}$
to each of the $A$ and $B$ variables. Since
this equation is equivalent to the
group cocycle condition, this $B\delta A$ theory is presumably
a representation of a different inequivalent 3-cocycle in the
$Z_{p}\times Z_{p}$
Dijkgraaf-Witten model. This is interesting since normally in gauge
theory the only possibility when writing down an action for a model
based on a direct product group is to take a sum of terms, one
for each factor, as in (\ref{cs2}).

\section{DW Models in Higher Dimensions}

An immediate question at this point is whether the higher dimensional
extensions of the Dijkgraaf-Witten model can also be interpreted within
the formalism we have been discussing. In $n=2 m$ dimensions, the
action one would take, based on a single 1-colour field, is clearly
a $\cup$-product of $m$ copies of $\delta A$. In terms of the
Boltzmann weight, one has
\bea
W[\sigma] = \exp\{\beta\, <\delta A \cup \cdots \cup \delta A,
\sigma >\} \;\;.\label{a}
\eea
Since this structure is a ``total derivative'', the Boltzmann weight is
always 1 on a closed $2m$-manifold, and no interesting phases can
result. While the group cohomology of $Z_{p}$ with $U(1)$ coefficients
is trivial in even dimensions, this is
not so with $Z_{p}$ coefficients. In fact, a simple application
of the universal coefficient theorem \cite{Rot},
\be
H^{n}(X,G) = H^{n}(X,Z) \otimes G \; \oplus \; Tor(H^{n+1}(X,Z),G)\;\;
\ee
to the result $H^{even}(BZ_{p},Z) = Z_{p}$ and  $H^{odd}(BZ_{p},Z) = 0$,
shows that
\be
H^{n}(BZ_{p},Z_{p}) = Z_{p}
\ee
for all nonnegative $n$.
In particular for $n=4$, when the flatness condition is imposed
and $s^{p^{3}} = 1$,
eqn. (\ref{a}) provides a representation of the $4$-cocycle. In this
particular model, the trouble is that when one multiplies together all
the $W$ factors for a closed complex, the total Boltzmann weight is $1$.
Since the Boltzmann weights are
actually $Z_{p}$ valued, it would be fascinating if they
could be realized in some other lattice model in even dimension.

For $2m+1$ dimensions, one easily writes the higher dimensional analogue
of the 3d Chern-Simons term. One takes the Boltzmann weight,
\bea
W[\sigma] = \exp\{\beta \, < A\cup \delta A \cdots \cup \delta A,
\sigma > \}\;\; ,
\eea
where one has $m$ factors of $\delta A$ in the action. The same analysis
that we have given earlier goes through without difficulty, and one
finds a subdivision invariant model when the factor $s = e^{\beta}$
is a $p^{m+1}$-root of unity. These would be concrete realizations of
the more abstract models implicit in \cite{DW}.

We also remark that, as in three dimensions, we have the freedom
to consider the $2m + 1$ dimensional model, with an array of different
$1$-colour fields. For example, in five dimensions, we obviously can
define models with the following Boltzmann weights,
\bea
W[\sigma] = \exp \{\beta <B^{(1)} \,\cup \, \delta A^{(1)} \,\cup \,
\delta A^{(1)}, \sigma>\}\;\;,\non\\
W[\sigma] = \exp \{\beta <B^{(1)} \,\cup \, \delta B^{(1)} \,\cup \,
\delta A^{(1)}, \sigma>\}\;\;,      \non\\
W[\sigma] = \exp \{\beta <A^{(1)} \,\cup \, \delta B^{(1)} \,\cup \,
\delta C^{(1)}, \sigma>\}\;\;.
\eea
The expectation would be that such models are related in some way
to the single $1$-colour model for product groups.

\section{General Models}

Let us now attend to the description of some potentially interesting
new models in higher dimensions.
In particular, we begin by considering a
four-dimensional theory, which involves the new feature of a $2$-colour
field.
The Boltzmann weight of an ordered $4$-simplex $[0,1,2,3,4]$ is defined by:
\be
W[[0,1,2,3,4]] = \exp\{\beta <B^{(2)} \,\cup \,
\delta A^{(1)}, [0,1,2,3,4]>\} \;\;,
\ee
where $B^{(2)}$ and $A^{(1)}$ are $2$- and $1$-colour fields, respectively.

To analyze the subdivision properties of this model, we consider
the introduction of a new vertex $c$ at the centre of the simplex
$[0,1,2,3,4]$, which is then joined to all the vertices, namely
\bea
[0,1,2,3,4] &\rightarrow& [c,1,2,3,4] - [c,0,2,3,4] + [c,0,1,3,4]\non\\
&-&[c,0,1,2,4] + [c,0,1,2,3]     \;\;.
\eea
As before, we declare this new vertex to be the first in the total
ordering of all vertices. It follows immediately that
\bea
& &W[[0,1,2,3,4]] s^{-<\delta B \,\cup\, \delta A, \, [c,0,1,2,3,4]>}
= W[[c,1,2,3,4]]\\
& &W[[c,0,2,3,4]]^{-1}\, W[[c,0,1,3,4]]\,
W[[c,0,1,2,4]]^{-1}\, W[[c,0,1,2,3]]\;\;.\non
\eea
Subdivision invariance of this
four dimensional theory is now guaranteed by imposing quantization of
the coupling $s^{p^{2}} = 1$, as well as
a restriction of the colourings to those satisfying the flatness conditions
\bea
[ \delta B^{(2)} ] = [ \delta A^{(1)}] = 0\;\; .\label{ba4d}
\eea
The subdivision invariant Boltzmann weight is then:
\be
W[[0,1,2,3,4]] = \exp\{\frac{2 \pi i k}{p^{2}}\,
B_{012}\,( A_{23}\,+\, A_{34} \,-\,
[ A_{23} + A_{34}] )\}\;\;.
\ee

The above subdivision move
is known as an Alexander move of type $4$ \cite{Alex}, or equivalently
a $(1,5)$ move \cite{Gross}. In order to complete the proof of
the combinatorial invariance of this four dimensional theory,
we also need to check invariance under a remaining set of subdivision moves.
These are conveniently represented by a set of $(k,l)$ moves,
where $(k+l = 6, k=1,\cdots,5)$ \cite{Gross}.  However,
it is straightforward to check invariance under the
remaining moves, by an analysis similar to the above.

We mention also that the $2$-colour field enjoys a gauge invariance
of the form:
\be
B'_{012} = [B_{012} - L_{01} - L_{12} + L_{02}]\;\;,
\ee
where $L$ is a $1$-colour field. As in the previous models, the
Boltzmann weight is gauge invariant for closed complexes.

In this four dimensional example, there is hope of finding a nontrivial
phase in the Boltzmann weight when one has solutions to (\ref{ba4d}) which
do not reduce to solutions in the strong sense, when the mod-p brackets
are removed. Experience in the 3d DW theory suggests that one find a
4d example where torsion is present in both $H_{1}(K,Z)$ and $H_{2}(K,Z)$.
The manifold $RP^{3}\times S^{1}$ fills the bill, and it will be
interesting to do an explicit computation of that partition function.

Moving on, we can now identify a series of models in $n$ dimensions
with Boltzmann weight given by:
\be
W[\sigma] = \exp\{\beta <B^{(r)}\,\cup \, \delta A^{(n-r-1)}, \sigma >\}
\;\;,
\ee
where $\sigma = [0,1,\cdots, n]$ is an $n$-simplex. In this case, the
colour degrees are $r$ and $(n-r-1)$ respectively, and again
subdivision invariance requires that $s = e^{\beta}$ is a $p^{2}$
root of unity, with field configurations being restricted by the
flatness conditions:
\bea
[\delta B^{(r)} ] = [\delta A^{(n-r-1)} ] = 0 \;\; .
\eea

At this point, it is worth remarking that non-trivial solutions
to these flat conditions will generically exist,
and these are
enumerated by the relevant cohomology groups, $H^{r}(K,Z_{p})$
and $H^{(n-r-1)}(K,Z_{p})$, of the complex $K$.

In $2m + 1$ dimensions, we can construct models with Boltzmann weight
\be
W[\sigma] = \exp\{\beta <B^{(m)} \,\cup\, \delta B^{(m)}, \sigma > \}\;\;,
\ee
or
\be
W[\sigma] = \exp\{\beta <B^{(m)} \,\cup\, \delta A^{(m)}, \sigma > \}\;\;,
\ee
where $\sigma = [0,1,\cdots , 2m+1]$, and $B^{(m)}$ and $A^{(m)}$
are independent $m$-colour fields, which, as usual, will be
restricted by the relevant flatness condition.
The important point to note here is that these models have a structure
distinct from the higher-dimensional Chern-Simons type theories
of the previous section, which were based only on $1$-colour fields.

It is also possible to consider extensions of these models in which the
$B$ and $A$ fields take values in different groups,
$Z_{p}$ and $Z_{q}$, say,
and with the scale parameter being chosen as $s^{pq} =1$.

To conclude this survey of models, we remark that theories which
include a $0$-colour field lead to a trivial Boltzmann
weight. The reason for this can be seen most easily in two
dimensions. Taking the Boltzmann weight of the simplex $[0,1,2]$ to be:
\be
W[[0,1,2]] = \exp\{\beta < B^{(0)} \,\cup \, \delta A^{(1)}, [0,1,2]>\}\;\;,
\ee
we
one finds that the relevant flatness conditions are
\be
[\delta B^{(0)}] = 0\;\;, [\delta A^{(1)}] = 0 \;\;.
\ee
However, the inequivalent solutions to the $0$-colour flatness
condition impose the constraint that the $B^{(0)}$ field is constant
on the vertices. One then sees that the Boltzmann weight is a ``total
derivative", and will always be $1$ on a closed $2$-manifold.

\section{Concluding Remarks}

It is clear that when the scale parameter $s=1$
the theories described above reduce simply to a sum over all
gauge inequivalent solutions to the flatness conditions.
Such an invariant is itself non-trivial, and thus the even dimensional
models presented above certainly differ from the Franz-Reidemeister
torsion, which is trivial in those dimensions.
Our main interest, of course, is in obtaining more subtle
behaviour at the non-trivial roots of unity.
One should note that in all the theories described, the central identity
obtained involves a product of $(n + 2)$ factors of the Boltzmann weight.
In \cite{top}, a variation of the cup product was used to define
a subdivision invariant lattice model in four dimensions.
In that case, a similar
identity involving six factors of the Boltzmann weight allowed one
to establish triviality of the invariant. The reason for this is that
the model was defined with an assignment of arbitrary group elements
to each link, without the imposition of flatness restrictions.
Perhaps, it is worth mentioning the possibility that expectation
values of gauge invariant observables, beyond the partition function,
may also yield some interesting structures, but we leave that for
the future.

{\bf Acknowledgements}\\
D.B. would like to thank R. Dijkgraaf, and M. de Wild Propitius, for
discussions.


\begin{thebibliography}{99}
\bibitem{DW} R. Dijkgraaf and E. Witten, {\em Topological Gauge Theories
and Group Cohomology}, Commun. Math. Phys. 129 (1990) 393.
\bibitem{AS} A.S. Schwarz, {\em The Partition Function of a Degenerate
Quadratic Functional and the Ray-Singer Invariants}, Lett. Math.
Phys. 2 (1978) 247.
\bibitem{JM} J. Munkres, {\em Elements of Algebraic Topology},
Addison-Wesley, Menlo Park, 1984.
\bibitem{Rot} J. Rotman, {\em An Introduction to Algebraic Topology},
Springer-Verlag, New York, 1988.
\bibitem{Still} J. Stillwell, {\em Classical Topology and Combinatorial
Group Theory}, Springer-Verlag, New York, 1980.
\bibitem{AC1} D. Altschuler and A. Coste, {\em Quasi-Quantum Groups,
Knots, Three-Manifolds, and Topological Field Theory}, Commun. Math. Phys.
150 (1992) 83.
\bibitem{AC2} D. Altschuler and A. Coste, {\em Invariants of
Three-Manifolds from Finite
Group Cohomology}, J. Geom. Phys. 11 (1993) 191.
\bibitem{FQ} D.S. Freed and F. Quinn, {\em Chern-Simons Theory
with Finite Gauge Group}, Commun. Math. Phys. 156 (1993) 435.
\bibitem{tric} W. Kuhnel, {\em Triangulations of Manifolds with
Few Vertices}, in {\em Advances in Differential Geometry and
Topology}, ed. F. Tricerri, World Scientific, 1990, pg. 59.
\bibitem{Alex} J.W. Alexander, {\em The Combinatorial Theory of
Complexes}, Ann. Math. 31 (1930) 292.
\bibitem{Gross} M. Gross and S. Varsted, {\em Elementary Moves and
Ergodicity in d-Dimensional Simplicial Quantum Gravity},
Nucl. Phys. B378 (1992) 367.
\bibitem{top} D. Birmingham and M. Rakowski, {\em Combinatorial
Invariants from Four Dimensional Lattice Models}, Int. J. Mod. Phys. A.,
to appear.
\end{thebibliography}
\end{document}